\documentclass[aps,preprint]{revtex4}
\usepackage[dvips]{graphicx}
\usepackage{amssymb}
\usepackage{amsmath}
\usepackage{color}

\usepackage[normalem]{ulem}
\newcommand{\soutr}{\bgroup\markoverwith{\textcolor{red}{\rule[.5ex]{2pt}{1pt}}}\ULon}

\usepackage{cancel}

\begin{document}
\draft

\title{Hearing Euler characteristic of graphs}

\author{Micha{\l} {\L}awniczak,$^{1}$ Pavel Kurasov,$^{2}$ Szymon Bauch,$^{1}$ Ma{\l}gorzata Bia{\l}ous,$^{1}$  Vitalii Yunko,$^{1}$ and Leszek Sirko$^{1}$}
\address{$^{1}$Institute of Physics, Polish Academy of Sciences, Aleja  Lotnik\'{o}w 32/46, 02-668 Warszawa, Poland\\
$^{2}$Department of Mathematics, Stockholm University, S-106 91 Stockholm, Sweden\\
}
\date{\today}

\begin{abstract}
The Euler characteristic $\chi =|V|-|E|$ and the total length $\mathcal{L}$ are the most important topological and geometrical  characteristics  of a metric  graph. Here,  $|V|$ and $|E|$ denote the number of vertices and edges of a graph.
 The Euler characteristic determines the number $\beta$ of independent cycles in a graph
while the total length determines the asymptotic behavior of the energy eigenvalues via the Weyl's
law. We show theoretically and confirm experimentally that the Euler characteristic can be determined (heard) from a finite sequence of the lowest eigenenergies $\lambda_1, \ldots, \lambda_N$ of a simple
quantum graph, without any need to inspect the system visually. In the experiment quantum graphs are simulated by microwave networks.  We demonstrate that the sequence of the lowest resonances of microwave networks with $\beta \leq 3$ can be directly used in  determining whether a network is planar, i.e., can be embedded in the plane.  Moreover, we show that the measured Euler characteristic $\chi$  can be used as a sensitive revealer of the fully connected graphs.

\end{abstract}


\bigskip
\maketitle

\section{Introduction}

The problem of seven bridges of K\"onigsberg considered by Leonhard Euler in 1736  \cite{Euler1736} was one of the most notable mathematical
achievements which laid the foundations of graph theory and topology.
In 1936 this seeding idea was used by Linus Pauling in physics \cite{Pauling36}
in order to describe a quantum particle moving in a physical network, the model known today as a quantum graph.

The idea of quantum graphs was further extensively developed in Refs. \cite{Exner88,Kottos1997, Blumel2002, BK, Pluhar2014}.
 In the considered model a metric graph  $\Gamma=(V,E)$ is formed by the edges $e \in E$ connected together at the vertices
$v \in  V$.
Each edge is seen as an interval on (a separate copy of) the real line $\mathbb{R}$
having the length $l_e$, then the vertices can be defined as disjoint unions of edge endpoints.
Let us consider the Laplace operator $L(\Gamma) = -\frac{d^2}{dx^2}$  acting in the Hilbert space of square
integrable functions on $\Gamma$ satisfying in addition the standard vertex conditions (also called natural, Neumann or Kirchhoff): the function is continuous at the vertex $v$ and
the sum of oriented derivatives at the vertex $v$ is equal to zero.
Such Laplacian is uniquely determined by the metric graph, is self-adjoint and
its spectrum is pure discrete \cite{BK}. Moreover, the operator is non-negative with zero
being a simple eigenvalue (provided the graph is connected) with the eigenfunction
given by the constant.
 For more details on quantum graphs, we can refer the reader to the book \cite{BK} and the references therein.
Quantum graphs were used  to simulate, e.g., mesoscopic quantum systems \cite{Kowal1990,Imry1996}, quantum wires \cite{Sanchez1988}, and optical waveguides \cite{Mittra1971}.

In this letter we report breakthrough results on topology of quantum graphs and microwave networks.  We show
that measuring several dozen of eigenvalues of the system one may recover its Euler characteristic without seeing a graph, i.e., knowing the number of the graph's vertices and edges. In particular one may even determine structural properties of the network, e.g., whether the
graph is planar or fully connected.

The original formula for $\chi $  \cite{Ku05} requires knowledge of all eigenenergies
of the system
and plays a very important role in the study of inverse problems for quantum graphs, but its applicability to
laboratory measurements is limited, since only a finite number of eigenenergies can be obtained in any
real world experiment.

From the experimental point of view it is important to point out that quantum graphs can be modeled by microwave networks \cite{Hul2004,Lawniczak2008, Hul2012,Sirko2016,Dietz2017,Lawniczak2019b}.  It is attainable because both systems are described by the same equations: the one-dimensional Schr\"odinger equation  appearing in  quantum graphs is formally equivalent to the telegrapher's equation for  microwave networks \cite{Hul2004,Sirko2016}. Microwave networks, as the only ones, allow for the experimental simulation of quantum systems corresponding to all three classical ensembles in the  random-matrix theory (RMT): the systems with $T$ invariance belonging to Gaussian orthogonal ensemble (GOE) \cite{Hul2004,Lawniczak2008,Hul2012,Dietz2017,Lawniczak2019} and Gaussian symplectic ensemble (GSE) \cite{Stockmann2016}, and the systems without  $T$  invariance belonging to Gaussian unitary ensemble (GUE) \cite{Hul2004,Lawniczak2010,Allgaier2014,Bialous2016,Lawniczak2017,Lawniczak2019b}.

Microwave networks were successfully used, e.g., to demonstrate the usefulness of missing level statistics in variety of applications \cite{Bialous2016} and to show that there exist graphs which do not obey a standard Weyl's law, called non-Weyl graphs \cite{Lawniczak2019}.

The most important characteristics of a metric graph $\Gamma=(V,E)$ are the Euler characteristic $\chi =|V|-|E|$ and the total length $\mathcal{L} =\sum_{e\in E} l_e$.
The
Euler characteristic $\chi$ determines the number $\beta$ of independent cycles in a graph
\begin{equation}
  \beta = |E|-|V|+1 \equiv 1-\chi \,,\label{eq:cycles}
\end{equation}
while the total length $\mathcal{L}$ determines the asymptotics of a graph's eigenvalues $\lambda_n$ via the  Weyl's
formula
\begin{equation}
  \lambda_n = \Big(\frac{\pi}{\mathcal{L}}\Big)^2n^2 + \mathcal{O}(n) \,,\label{eq:Weyl}
\end{equation}
where $\mathcal{O}(n)$ is a function which in the limit $n\rightarrow +\infty $ is bounded by a constant.
The number of independent cycles measures how different a graph
is from a tree and is equal to the number of edges that have to be deleted
to turn the graph into a tree.

It might seem that the determination of both characteristics would require the knowledge of
the whole sequence of eigenvalues. Such an assumption is natural in mathematics and allows to derive the precise formulas for
$ \mathcal{L} = \pi\lim_{{\textmd n\rightarrow +\infty }}\frac{n}{k_n} $ and $\chi = X(t) \vert_{t \geq t_0}$ \cite{Ku05,Ku06}, where

 \begin{equation}
  X(t) := 2 + 2\pi \sum_{n=1}^{ \infty} \cos(k_n/2t)\Big(\frac{\sin(k_n/4t)}{k_n/4t}\Big)^2 , \label{eq:chi}
\end{equation}
 and $k_n$ are the square roots of the eigenenergies  $\lambda_n$ and $t_0=\frac{1}{2l_{min}}$ with  $l_{min}$ being the length of the shortest edge of a simple graph.
While derivation of formula for $ \mathcal{L} $ is elementary, formula (\ref{eq:chi}) can be obtained either from the trace formula   \cite{GuSm3,KuNo8,Ro11}  connecting the spectrum to the set of periodic
orbits on $\Gamma$  \cite{Ku05}  or by analyzing the heat kernels \cite{Ro11}.

The knowledge
of the whole spectrum allows one to reconstruct the metric graph, provided the edge
lengths are rationally independent (see e.g. \cite{vBe1,GuSm3,KuNo8}) thus providing an affirmative
answer to the classical question asked by Mark Kac \cite{Kac4} adopted to quantum graphs as
``Can one hear the shape of a graph?" \cite{Hul2012}.

However, in the real world experiments there is no chance to determine the entire spectrum. For example in microwave networks because of openness of the systems and the existence of internal absorption one can measure up to several hundreds of eigenfrequencies. Moreover, one cannot guarantee that the edge lengths are
rationally independent, therefore it is natural to investigate the question whether
the total length  $\mathcal{L}$ and the Euler characteristic  $\chi$ can be reconstructed directly from the
spectrum without determining a precise form of the graph. Formulas for $\mathcal{L}$ and $X(t)$
provide such a possibility but their character is completely different.
The total length $\mathcal{L}$ is a positive real number, hence to determine it with a high precision one needs to know high energy eigenvalues $\lambda_n$. More eigenvalues are determined the better approximation of $\mathcal{L}$ is obtained.
The Euler characteristic $\chi$ is an integer number (often negative), hence to
determine it precisely it is enough to know the right hand side of (\ref{eq:chi})
 with an error less than 1/2.
Therefore, knowing that in the experiment only a limited number of the eigenvalues can be measured,  we shall concentrate in this letter on determining the Euler characteristic $\chi$.

\section{A new formula for the Euler characteristic}

The series in formula (\ref{eq:chi}) for the Euler characteristic is slowly converging. Its application requires the measurements of several hundreds or even more of eigenenergies which in the most cases is not achievable. Therefore, we derived a new function

\begin{equation}
 X(t) := 2 + 8\pi^2\sum_{k_n\neq 0}\frac{\sin(k_n/t)}{(k_n/t)[(2\pi)^2-(k_n/t)^2]} , \label{eq:chi_2}
\end{equation}
which gives the Euler characteristic $\chi = X(t) \vert_{t \geq t_0}$  and is characterized by a much better convergence.
The details of derivation are given in the Appendix.

\section{Experimental implementation}

Let us assume that in the experiment the K lowest resonances (eigenvalues) are measured. We shall calculate the Euler characteristic
  $ \chi $ by
evaluating the function $ X(t)$  by substituting the infinite series with a finite sum   and  assuming that $t \geq t_0$.  Let us introduce
the function     $X_K(t)$ corresponding to a new formula (\ref{eq:chi_2})

\begin{equation}
  X_K(t)  = 2 + 8\pi^2\sum_{n=1}^K\frac{\sin(k_n/t)}{(k_n/t)[(2\pi)^2-(k_n/t)^2]} \,\,. \label{eq:c2}
\end{equation}

We are going to
analyze whether this function gives a good approximation for the Euler characteristic $\chi$  when $t = t_0 =\frac{1}{2l_{min}}$.
Comparing (\ref{eq:chi_2}) with (\ref{eq:c2}) we obtain $\epsilon = |X(t_0)  -X_K(t_0)|$. In order to guarantee the difference $\epsilon$ is less than 1/2, e.g. 1/4,  it is enough to take the first $K$ eigenvalues evaluated by the following formula

\begin{equation}
K  \simeq  |V|-1 + 2\mathcal L t_0 \left [1-\exp\left (\frac{-\epsilon \pi}{\mathcal L t_0}\right )\right ]^{-1/2}  \,. \label{eq:k2}
\end{equation}
 The details of the proof are given in the Appendix.

 The new formula for the Euler characteristic (\ref{eq:chi_2}) was tested experimentally using  planar and non-planar   microwave networks for which  the counting function of the number of resonances satisfies the Weyl's law \cite{Lawniczak2019}. For such networks the Euler characteristic is the same as for the  corresponding closed quantum graphs.

In Fig.~1(a) and (b) we present the schemes of  a planar  quantum graph $\Gamma$ with $|V|=4$ vertices and $|E|=6$ edges  and  a planar microwave network with the same topology as $\Gamma$. The total optical length of the microwave network is $\mathcal L=1.494\pm0.006$ m and the optical length of the shortest edge is $l_{min}=0.155\pm0.001$ m. The optical lengths $l^{opt}_i$ of the edges of the network  are connected with their physical lengths $l^{ph}_i$ through the relationship $l^{opt}_i = \sqrt{\varepsilon }l^{ph}_i$, where $\varepsilon=2.06$ is the dielectric constant of the Teflon used for the construction of the microwave cables. The quantum graph is a closed dissipationless system for which according to the definition of the Euler characteristic $\chi=|V|-|E|=-2$. One should point out that the lack of dissipation  is a standard  assumption considered in the mathematical analysis of graphs.

In Fig.~2(a) we show that the formula (\ref{eq:c2}) can be easily used to reconstruct the Euler characteristic of the microwave network in Fig.~1(b) and obtain the correct result $ \chi = -2$.
As all real life systems, this system is open and is characterized by small dissipation  \cite{Lawniczak2010}.
 The resonances $\nu_1, \ldots, \nu_N$  of the microwave network required for the evaluation of the Euler characteristic  were determined from the measurements of a one-port scattering matrix $S(\nu)$ of the network using the vector network analyzer (VNA) Agilent E8364B.

 One should note that it is customary for microwave systems to make measurements of the scattering matrices  as  a function of microwave frequency $\nu$. Then  the real parts of the wave numbers  $k_n$ are directly related to the positions $\nu_n$ of the resonances  $\mathrm{Re\,}k_n=\frac{2\pi }{c}\nu_n$. The VNA was connected to the microwave network with the flexible HP 85133-616 microwave cable which is equivalent to attaching an infinite lead to a quantum graph \cite{Lawniczak2019}. Before each measurement the VNA was calibrated using the Agilent 4691-60004 electronic calibration module to eliminate the errors in the measurements.

 In order to avoid the missing resonances we analyzed the fluctuating part of the integrated spectral counting function $N_{fl}(\nu_i)=N(\nu_i)-N_{av}(\nu_i)$ \cite{Dietz2017}, that
is the difference of the number of identified eigenfrequencies
$N(\nu_i) = i$  for ordered frequencies $\nu_1 \leq \nu_2 \leq \ldots$ and
the average number of eigenfrequencies $N_{av}(\nu_i)$ calculated in the considered frequency range. Using this well known method \cite{Dietz2017} we were able to identify the first $N=106$ resonances in the frequency range of $0.31-10.76$ GHz.  The problem with the resolution of the resonances begins for $N \simeq 100-150$ but then the sensitivity of the Euler characteristic  (\ref{eq:chi_2}) for the missing resonances is very weak.

 In Fig.~2(a) we show the approximation function for the Euler characteristic $X_K(t)$ (\ref{eq:c2}) calculated using the first $K=28$ (green full line) and $K=106$ (red dash-dotted line) experimentally measured resonances of the system, respectively. The value $K=28$ was estimated from the formula (\ref{eq:k2}) assuming that $\epsilon = 1/4$ and  taking into account the optical size of the network $\mathcal{L}t_0=4.82\pm0.05$. In Fig.~1(f)
 we show, as an example, the modulus of the scattering matrix  $|S(\nu)|$  of the experimentally studied  microwave network  $\Gamma$ with $|V|=4$ measured in the frequency range $\nu=3.0-4.5$ GHz.
 Fig.~2(a) demonstrates that it is enough to use the first $K=28$ resonances (green full line) to identify a clear plateau close to the expected value $\chi=-2$. This plateau extends  from $3\textrm{ m}^{-1}<t<6\textrm{ m}^{-1}$ and includes the parameter $t_0 \simeq 3.23$ $\textrm{ m}^{-1}$ which was used for the evaluation of the  required number of resonances $K=28$ (see the formula (\ref{eq:k2})). The Euler characteristic calculated for  $K=N=106$ resonances (red dash-dotted line) displays a very long plateau along the expected value $\chi=-2$. The plateau extends from  $3~\textrm{m}^{-1}<t<17\textrm{ m}^{-1}$ showing that we actually deal with the excessive number of resonances  required for the practical evaluation of the Euler characteristic. Just for the comparison we also show in Fig.~2(a) the Euler characteristic calculated from the Eq.~(\ref{eq:chi}) using the first  $K=28$ resonances (brown dotted line). As expected, the  formula (\ref{eq:chi}) shows much worse convergence to the predicted value of $\chi=-2$.

Although for the analysis of the convergence of the formula (\ref{eq:chi_2}) (see the Eq. (\ref{eq:k2})) we used the graph's parameters $\mathcal L$ and $t_0$ in the real applications we do not need them. The power of the formula (\ref{eq:chi_2}) stems from the fact that the sequence of the lowest resonances allows for the determination of the Euler characteristic without knowing physical parameters of the graph. In practice, if a plateau in $X_K(t)$ along a given integer number is not observed it means that the number of resonances used in the calculations is
insufficient  and it ought to be increased.

It is important to point out that  the formula (\ref{eq:chi_2}) allows also for the determination whether a system is planar.
In the analyzed cases  of the graph $\Gamma$    and the microwave network the number of cycles yielded from the formula (\ref{eq:cycles})
is $\beta=1-\chi=3$. In accordance with the Kuratowski's theorem \cite{Kuratowski1930} every non-planar graph should contain $K_5$  (the complete graph on $5$ vertices)
or $K_{3,3}$  (the complete bipartite graph on $3$ and $3$ vertices) as
 subgraphs.  These graphs have $6$ and $4$ cycles, respectively, therefore, without even seeing a graph or having a complete information  about the number of vertices and edges we found out that the graph is planar and the microwave network simulates the planar graph.

Let us now analyze the situation of non-planar fully connected (complete in the mathematical terminology) graphs and
networks. In Fig.~1(c)~and~(d)
we present  the non-planar fully connected  quantum graph  $K_5$, complete graph on $ |V|=5 $ vertices, characterized
by the Euler characteristic $\chi=-5$,  and  the microwave network with the same topology.  The total optical length of the microwave network is $\mathcal L=3.949\pm0.010$ m and the optical length of the shortest edge is $l_{min}=0.202\pm0.001$ m. To perform the measurements of the first $N=132$ eigenresonances the network was connected to the VNA  with the flexible microwave cable (see Fig.~1(e)). In Fig.~1(f) we show the modulus of the scattering matrix  $|S(\nu)|$  of this network  ($|V|=5$) measured in the frequency range $\nu=3.0-4.5$ GHz.

 The approximation function for the Euler characteristic  (\ref{eq:c2}) calculated for the first $K=74$ (green full line) and $K=132$ (red dash-dotted line) experimentally measured resonances of the system, respectively, is shown in Fig.~2(b). The value $K=74$ was estimated from the formula (\ref{eq:k2}) assuming again that $\epsilon = 1/4$ and  taking into account the optical size of the  $ K_5 $ network $\mathcal{L}t_0 = 9.74 \pm 0.10$.
Fig.~2(b) shows that using  $K=74$ resonances measured for the non-planar  microwave network in Fig.~1(d) the correct Euler characteristic $\chi=-5$ can be easily evaluated (full green line). In this case a long plateau close to the expected value $\chi=-5$ is seen in the parameter range $2.5 \textrm{ m}^{-1}<t<4 \textrm{ m}^{-1}$. The situation improves even further for the Euler characteristic calculated for $K=N=132$ resonances measured in the frequency range of $0.19-5.12$ GHz (red dash-dotted line). In this case the plateau is extended in the range  $2.5\textrm{m}^{-1}<t<7.5\textrm{m}^{-1}$  clearly indicating that the Euler characteristic can be also properly evaluated using much less resonances.
In Fig.~2(b) we also show the approximation function for the Euler characteristic $X_K(t)$ calculated from the Eq.~(\ref{eq:chi}) using the first  $K=74$ resonances (brown dotted line). It is visible that the formula (\ref{eq:chi}) yields the results which are far away from the predicted value of $\chi=-5$ and can be only used for much higher number of resonances $K=1243$ (see the formula \eqref{est4} in the Appendix).

For the analyzed graph $K_5 $  and the  microwave network the number of cycles calculated from  the formula (\ref{eq:cycles})
is $\beta=1-\chi=6$. Since the number of cycles is higher than $3$ we cannot directly assess whether the system is planar or not since the application of the Kuratowski's theorem requires the full information about the topology of the investigated graph which in principle is not available.
In such a situation, in general, we can only test whether graphs and networks analyzed by us are fully connected. The fully connected simple networks and graphs are especially interesting because there is an explicit link between the number of vertices $|V|$ of a graph and the Euler characteristic

\begin{equation}
|V|= \frac{3+\sqrt{9-8\chi}}{2} \,. \label{eq:fully}
\end{equation}
This formula holds for both planar and non-planar graphs.
Applying the formula (\ref{eq:fully}) in the case of the  microwave network   $\Gamma$  with the measured  Euler characteristic $\chi=-2$ we get $|V|=4$. Since the number of vertices yielded by the formula (\ref{eq:fully}) is integer it means that our planar network is also fully connected.   In the case of the network $K_5$  with the measured  Euler characteristic $\chi=-5$ we directly find out that the number of vertices of the network is $|V|=5$,  in obvious agreement with the number of the vertices  of the network. Therefore, in this case the experimental evaluation  of the Euler characteristic $\chi$  allowed us to find out that we deal with the fully connected non-planar  $K_5$ network.

In summary, we showed that the Euler characteristic $\chi$ can be determined (heard) from a finite sequence of the lowest resonances $\nu_1, \ldots, \nu_N$ of a microwave network. We also demonstrated that the spectrum of a simple microwave network can be used to find  the number $\beta=1-\chi $ of independent cycles. If $\beta \leq 3$ then a studied system is planar. Moreover, the  Euler characteristic $\chi$  allows to identify whether the networks and graphs are fully connected. In such cases it is possible to determine the number of vertices and edges of the systems. Thus, we clearly showed that the Euler characteristic $\chi$ is a powerful measure of graphs or networks properties, including topology,  complementing in the important way the inverse spectral methods that require the infinite number of eigenenergies or resonances for their application.

\section{Acknowledgements}
This work was supported in part by the National Science Centre, Poland, Grant No. 2016/23/B/ST2/03979,  the Swedish Research Council (Grant D0497301) and the Center for Interdisciplinary Research
(ZiF) in Bielefeld in the framework of the cooperation group on {\it Discrete
and continuous models in the theory of networks}.

\section{Appendix}

\subsection{A new formula for the Euler characteristic}

The formula \eqref{eq:chi} for the Euler characteristic
 derived in \cite{Ku05,Ku06} using the trace formula coming from \cite{Ro11,GuSm01,KuNo8} is not effective when the number of known eigenvalues is limited.

 Therefore, we derived a new formula for the Euler characteristic $\chi $ with a better convergence of the series.
 A new formula is obtained by applying the distribution $ u(k) $ \cite{Ku05,Ku06}

\begin{equation}
 u(k) := 2\delta(k) +\sum_{k_n>0}(\delta(k-k_n)+\delta(k+k_0)) = \chi \delta(k)+\frac{\mathcal{L}}{\pi} + \frac{1}{\pi}\sum_{p \in P}\ell(\mbox{prim}(p))S(p)\cos(k\ell(p)) \,,\label{eq:uk}
\end{equation}
where the sum is taken over all periodic orbits $P$ on $\Gamma$,  $ \ell(p) $ is the length of the orbit $ p $, and the coefficients $S(p)$ are products of scattering coefficients along the orbit $p$,
to the test function

\begin{equation}
\varphi (x) =
\left\{
\begin{array}{ll}
1 - \cos (2 \pi x), & 0 \leq x \leq 1; \\
0, & \mbox{otherwise},
\end{array} \right.
\end{equation}
which is continuous and has continuous first derivative.

The formula (\ref{eq:uk}) alone shows that
knowing the spectrum, equivalently, the distribution on the left hand side of the
formula (\ref{eq:uk}), allows one to reconstruct the Euler characteristic $\chi$.

The Fourier transform of the test function $\varphi (x)$ is
\begin{equation}
\sqrt{2\pi} \hat{\varphi} (k) =
\int_0^1 (1- \cos (2 \pi x)) e^{-ikx} dx
 = - i (e^{-ik} -1) \frac{4 \pi^2}{k[k^2 - (2\pi)^2]},
\end{equation}
and its real part is given by
\begin{equation}
\Re \sqrt{2\pi} \hat{\varphi} (k) =  - \frac{\sin (k)}{k}  \frac{4 \pi^2}{k^2- (2\pi)^2 }.
\end{equation}
The key point of the proof is to use the relation between the Fourier transforms of the distributions and the test functions
\begin{equation}
u [ \hat{\varphi}_t (k) ] = \hat{u} [ \varphi_t (x)].
\end{equation}
Applying $ \hat{u} $ to $ \varphi_t (x) $ for
 \begin{equation} \label{ineqt}
2t \ell_{\rm min} \geq 1 \Leftrightarrow t \geq t_0 = \frac{1}{2\ell_{\rm min}},
\end{equation}
where $ \ell_{\rm min} $ is the length of the shortest edge of the graph and therefore $ 2 \ell_{\rm min} $ is the length of the shortest periodic
orbit, we get
\begin{equation}
\hat{u} [\varphi_t(x)] = \chi, \quad t \geq t_0.
\end{equation}
Calculating $ u[ \hat{\varphi}_t(k)] $ we obtain a new formula \eqref{eq:chi_2} for the Euler characteristic
\begin{equation} \label{Euler2}
 \chi =  X(t) \vert_{t \geq t_0},
 \quad \; X(t)  = 2 + 8 \pi^2
\sum_{k_n \neq 0} \frac{\sin (k_n/t)}{(k_n/t) [(2\pi)^2-(k_n/t)^2] },
\end{equation}
improving the formula \eqref{eq:chi}.
The possible zeros in the denominator are not dangerous since they cancel with the zeroes in the numerator.

\subsection{The error estimate for the new formula}

We are interested in estimating how many resonances are needed to determine the Euler characteristic  $ \chi $, in other words how many terms  in the series are enough to evaluate $ X(t) $.
Since the Euler characteristic $ \chi $ takes integer values it is enough to require that the error $\epsilon$ is less than $ 1/2$:
\begin{equation}
 \epsilon = |X(t) -X_K(t)|_{\mid_{t=t_0}} = \left|8\pi^2\sum_{n=K+1}^{\infty}\frac{\sin(k_n/t_0)}{(k_n/t_0)[(2\pi)^2-(k_n/t_0)^2]}\right|  , \label{eq:c2chi}
\end{equation}
where
\begin{equation}
 X_K(t) = 2 + 8\pi^2\sum_{n=1}^K\frac{\sin(k_n/t)}{(k_n/t)[(2\pi)^2-(k_n/t)^2]} .  \label{eq:c2}
\end{equation}
Our claim is that it is enough to take
\begin{equation} \label{eq:k2}
 K > |V|-1 + 2\mathcal L t_0 \left [1-\exp\left (\frac{-\epsilon \pi}{\mathcal L t_0}\right )\right ]^{-1/2} ,
\end{equation}
where $\mathcal L t_0 = \frac{\mathcal L}{2l_{min}}$. For $\frac{\mathcal L}{2l_{min}}\gg 1$ the condition  (\ref{eq:k2}) can be approximated by
\begin{equation} \label{eq:k2app}
K>    |V|-1 + \frac{2}{\sqrt{\epsilon \pi}} \left ( \frac{\mathcal L}{2l_{min}} \right )^{3/2} .
\end{equation}

To prove (\ref{eq:k2}) we assume first that $ K $ is sufficiently large to guarantee that the denominator in  \eqref{eq:c2chi} is negative
$k_{K+1} > 2 \pi t_0. $
Taking into account the elementary lower estimate for the eigenvalues
\begin{equation} \label{estk}
 k_n^2 \geq \big( \frac{\pi}{\mathcal L} \big)^2 (n+1-|V|)^2, \end{equation}
where $ |V| $ is the number of vertices, we arrive at the following sufficient condition for the denominator to be negative:
\begin{equation} \label{estk2}
K > |V|-1 + \frac{\mathcal L}{\ell_{\rm min}}.
\end{equation}

Then the series can be estimated as
\begin{equation} \label{estk3}
\begin{array}{ccl}
\displaystyle \left\vert X(t_0)- X_K (t_0) \right\vert  & \leq &
\displaystyle  8\pi^2\sum_{n=K+1}^{\infty}\frac{\left|\sin(k_n/t_0)\right|}{(k_n/t_0)[(k_n/t_0)^2-(2\pi)^2]} \\[5mm]
& \leq & \displaystyle 8 \frac{(\mathcal L t_0)^3}{\pi} \sum_{n = K+1}^\infty \frac{1}{(n+1-|V|)[(n+1-|V|)^2 - 4 \mathcal L^2 t_0^2]}  \\[5mm]
& \leq & \displaystyle 8 \frac{(\mathcal L t_0)^3}{\pi} \int_K^\infty \frac{dx}{(x+1-|V|)[(x+1-|V|)^2 - 4 \mathcal L^2 t_0^2]}   \\[5mm]
& = & \displaystyle  \frac{\mathcal L t_0}{\pi}  \log\frac{(K+1-|V|)^2 }{(K+1-|V|)^2 - 4 \mathcal L^2 t_0^2} ,
\end{array}
\end{equation}
where we again used \eqref{estk} and substituted series with an integral on the last step.
Requiring that the error is less than $\epsilon$ leads to \eqref{eq:k2}.

\subsection{The error estimate for the original formula}

Using similar arguments we may derive a rigorous estimate
for the number of necessary resonances $K$ required in the case of the formula \eqref{eq:chi}
\begin{equation}
\label{est4}
 K > |V| -1 + \frac{32 \mathcal  L^2 }{\epsilon \pi^2} t_0^2 \equiv  |V| -1 + \frac{32}{\epsilon \pi^2} \Big( \frac{\mathcal  L }{2\ell_{\rm min}} \Big)^2 .
 \end{equation}
 Since the ratio $\frac{\mathcal  L }{2\ell_{\rm min}}$ in the formula (\ref{est4}) is raised to the second power the above estimate for $\frac{\mathcal L}{2l_{min}}\gg 1$ is definitely much worse than \eqref{eq:k2app}, which clearly explains why the old formula \eqref{eq:chi} for the Euler characteristic is ineffective in the real world applications.


\begin{figure}[tb]
\includegraphics[width=0.8\linewidth]{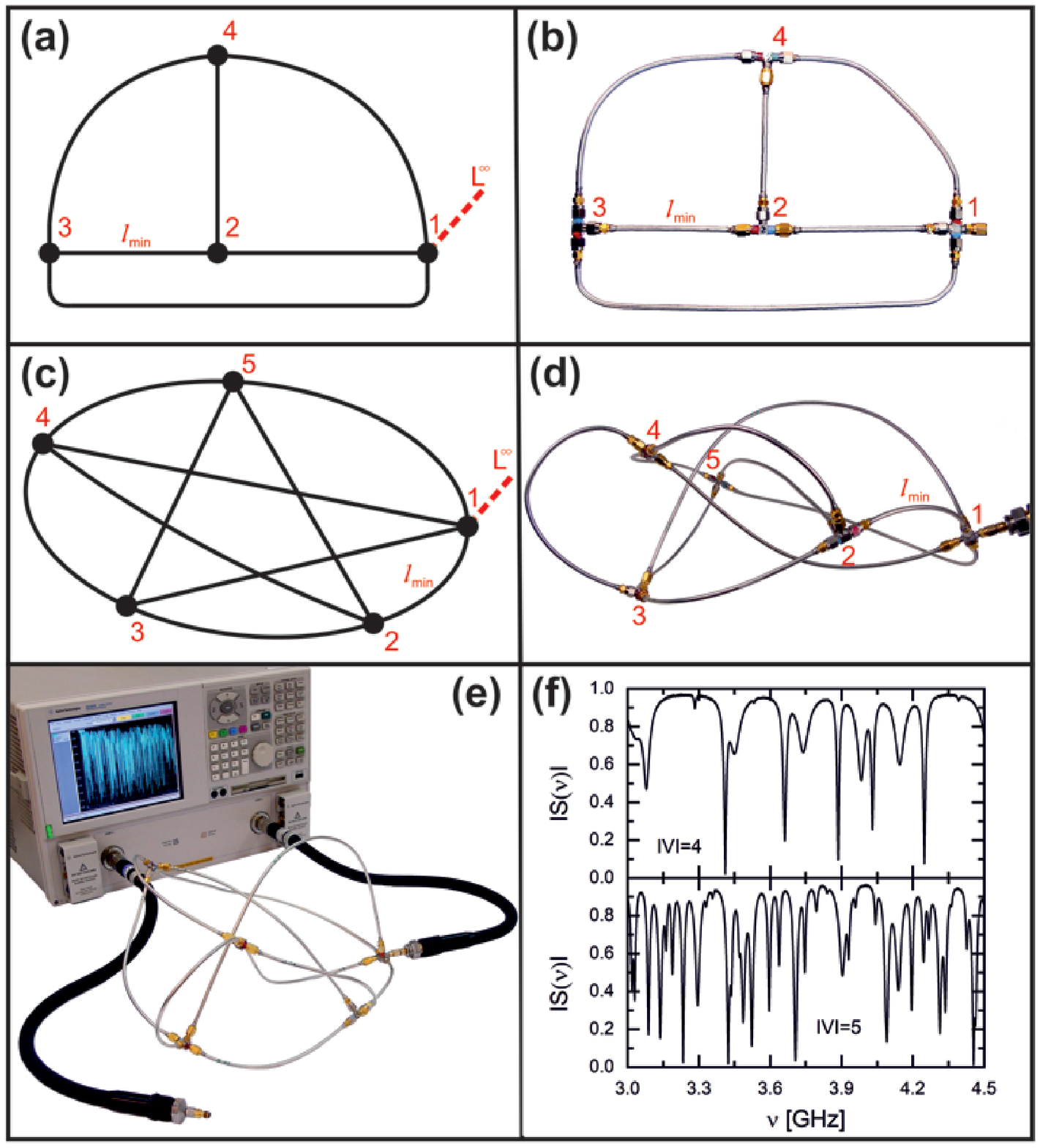}
\caption{
Panels (a) and (b) show the schemes of  a planar quantum graph with $|V|=4$ vertices and $|E|=6$ edges and a  microwave network  with the same topology. Panels (c) and (d) show the schemes of  a non-planar quantum graph with $|V|=5$ vertices and $|E|=10$ edges and a  microwave network  with the same topology.
 The microwave networks were connected to the vector network analyzer with the flexible microwave cable which is equivalent to attaching an infinite lead  to a quantum graph (panel (e)).
 Panel (f) shows the examples of the moduli of the scattering matrix  $|S(\nu)|$  of the  microwave networks with $|V|=4$  and $|V|=5$ vertices, respectively, measured in the frequency range $\nu=3.0-4.5$ GHz.
}
\label{Fig1}
\end{figure}

\begin{figure}[tb]
\includegraphics[width=1.0\linewidth]{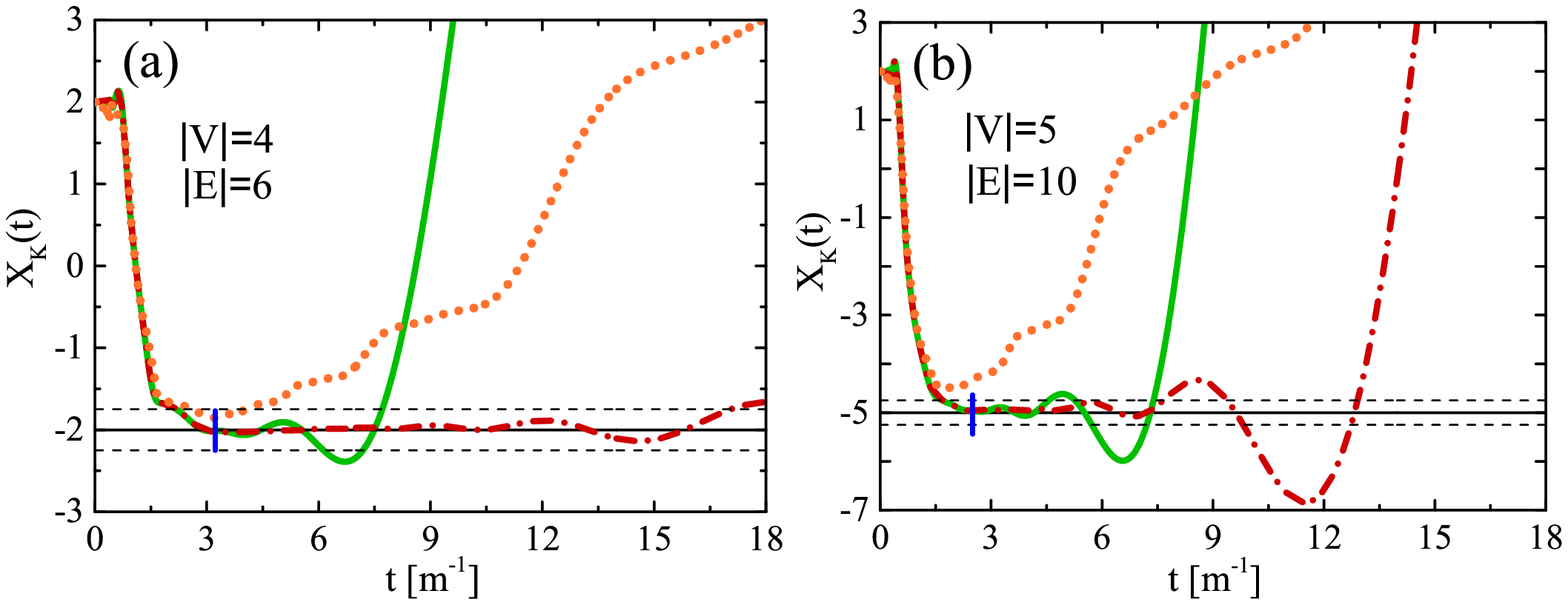}
\caption{
The approximation function for the Euler characteristic  $X_K(t)$ calculated for a planar microwave  network with  $|V|=4$ vertices and $|E|=6$ edges (panel (a)) and for  a non-planar fully connected microwave network with  $|V|=5$ vertices and $|E|=10$ edges (panel (b)).
The full green and red dash-dotted lines show the function $X_K(t)$  calculated from the Eq. (\ref{eq:c2}) for the first $K=28$ and $K=106$  resonances (panel (a)), and $K=74$ and $K=132$ resonances (panel (b)), respectively.
The blue vertical mark shows the value of $t_0=\frac{1}{2l_{min}}$ used for the evaluation of the required number of resonances $K=28$ (see the formula (\ref{eq:k2})) (panel (a)) and $K=74$ (panel (b)).
For the comparison we show the function $X_K(t)$ (brown dotted line) calculated from the Eq. (\ref{eq:chi}) using the first  $K=28$ (panel (a)) and $K=74$ (panel (b)) resonances.
The black full line shows the expected value of the Euler characteristic $\chi=-2$ (panel (a)) and  $\chi=-5$ (panel (b)). The black broken lines show the limits of the expected errors  $\chi \pm 1/4$.
}
\label{Fig2}
\end{figure}

\end{document}